\documentclass[reprint,aip,jcp,amsmath,citeautoscript,floatfix]{revtex4-1}

\usepackage{graphicx}
\usepackage{braket}
\usepackage{txfonts}
\usepackage{bm}
\usepackage{color}
\definecolor{myblue}{rgb}{0,0,1}

\usepackage[breaklinks=true,colorlinks=true,linkcolor=myblue,urlcolor=myblue,citecolor=myblue]{hyperref}

\newcommand{\dd}{\mathrm{d}}

\begin{document}

\title{Mixed quantum--classical modeling of exciton--phonon scattering in solids: Application to optical linewidths of monolayer MoS\texorpdfstring{\textsubscript{2}}{2}}
\author{Alex Krotz}
\author{Roel Tempelaar}
\email{roel.tempelaar@northwestern.edu}
\affiliation{Department of Chemistry, Northwestern University, 2145 Sheridan Road, Evanston, Illinois 60208, USA}

\begin{abstract}
We present a mixed quantum--classical framework for the microscopic and non-Markovian modeling of exciton--phonon scattering in solid-state materials, and apply it to calculate the optical linewidths of monolayer MoS$_2$. Within this framework, we combine reciprocal-space mixed quantum--classical dynamics with models for the quasiparticle band structure as well as the electron--hole and carrier--phonon interactions, parametrized against \textit{ab initio} calculations, although noting that a direct interfacing with \textit{ab initio} calculations is straightforward in principle. We introduce various parameters for truncating the Brillouin zone to select regions of interest. Variations of these parameters allow us to determine linewidths in the limit of asymptotic material sizes. Obtained asymptotic linewidths are found to agree favorably with experimental measurements across a range of temperatures. As such, our framework establishes itself as a promising route towards unraveling the non-Markovian and microscopic principles governing the nonadiabatic dynamics of solids.
\end{abstract}

\maketitle

\section{Introduction}

Rapid advances in materials engineering are challenging the traditional paradigms of materials behaviors. In particular, while solids are commonly thought of as high-dielectric media, there is a growing class of materials where dielectric screening is weak as a result of reduced dimensionality \cite{mueller2018two}, porosity \cite{kshirsagar2021strongly}, and/or incorporation of organic constituents \cite{li2023low}. For semiconducting materials, this may lead to the formation of stable excitons, i.e., electron--hole pairs with sizeable binding energies. In the presence of direct bandgaps, such tightly-bound excitons will be strongly absorbing \cite{mueller2018two}. Reduced dielectric screening may also lead to sizable interactions between electronic carriers and vibrations of the polar lattice, i.e., phonons \cite{wright2016electron, li2021exciton}. Much remains to be learned about the dynamical interplay of strongly-interacting electrons, holes, and phonons within a materials platform, posing a demand for new theoretical methods that are microscopic and non-Markovian.

\begin{figure*}
    \centering
    \includegraphics{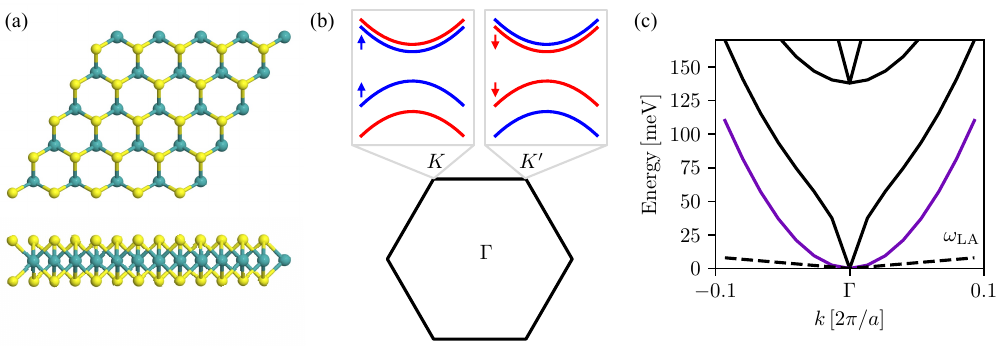}
    \caption{(a) Atomic structure of monolayer TMDs, with transition metals in teal and chalcogens in yellow. Shown are top view (upper) and side view (lower). (b) Hexagonal BZ with the high-symmetry points indicated. Shown on top is a schematic of the conduction and valence bands for spin up (blue) and spin down (red) near the $K$ and $K'$ points. (c) Calculated dispersions of the low-energy exciton states of MoS$_2$ (projected along the reciprocal $x$ direction). The lowest-energy exciton band shown in purple is included in our dynamics calculations. The energy of this exciton at the $\Gamma$ point is taken as a reference. Also shown is the dispersion of the LA phonon (dashed).}
    \label{fig:scheme}
\end{figure*}

Monolayer transition-metal dichalcogenides (TMDs) constitute an emerging class of direct-bandgap semiconducting materials \cite{mak2010AtomicallyThinMoS, splendiani2010emerging}, the atomic structure of which is shown in Fig.~\ref{fig:scheme} (a). In TMDs, weak dielectric screening emanates from reduced dimensionality \cite{chernikovExcitonBindingEnergy2014}. As a result, exciton binding energies in TMDs have been shown to range in the hundreds of meV \cite{chernikovExcitonBindingEnergy2014, latini2015excitons}, rendering these materials suitable for excitonic applications at room temperature. Higher-order carrier complexes have also been observed, including negatively-charged trions \cite{mak2013tightly}, consisting of two electrons interacting with one hole, as well as biexcitons \cite{you2015strong}. The technological interest in these materials is further motivated by their nontrivial topology. As a result of this topology, the band extrema where excitons form organize into spin-opposite degenerate ``valleys'' located at the inequivalent corners of the hexagonal Brillouin zone (BZ), as depicted in Fig.~\ref{fig:scheme} (b). Each valley is addressable through opposite circularly-polarized light \cite{xiao2012coupled, cao2012valley, kioseoglou2012valley, mak2012control, sallen2012robust, zeng2012valley}, opening up spintronic opportunities.

Carrier--phonon interactions in TMDs lead to a variety of excitonic scattering pathways, impacting optical lineshapes \cite{moody2015intrinsic, dey2016OpticalCoherenceAtomicMonolayer, christiansen2017phonon, li2021exciton}, photoluminescence enhancement \cite{chow2017phonon}, and valley polarizations \cite{yu2014valley}. Spectrally-resolved phonon sidebands observed for TMDs indicate carrier--phonon interactions to be sizeable and non-Markovian \cite{christiansen2017phonon, shree2018observation, li2021exciton}. Knorr and coworkers were able to model phonon sidebands by incorporating non-Markovian contributions into the semiconductor Bloch equations \cite{christiansen2017phonon}. Reiter and coworkers instead resorted to a time-convolutionless master equation, to similar effect \cite{lengersTheoryAbsorptionLine2020}. Recent work by Louie and coworkers accounted for the resulting absorption peak asymmetries by means of second-order many-body perturbation theory, which retained microscopic detail of the material \cite{chan2023exciton}. These studies are pushing the envelopes of theoretical modeling, offering much-needed insights into the unconventional behaviors of TMDs. However, a particular challenge in the microscopic modeling of TMDs is posed by rapid fluctuations of the dielectric function in the limit of small lattice momentum (wavevector) \cite{huser2013how}, which requires the BZ to be sampled at high resolutions \cite{qiu2015erratum, qiu2016screening}. Given the unfavorable computational cost scaling with increasing BZ resolution, it remains unclear to what extent existing theoretical methods are able to reach convergence while retaining a microscopic and non-Markovian incorporation of carrier--phonon interactions.
    
A comparatively inexpensive means to microscopically account for the non-Markovian interactions between electrons, holes, and vibrational modes is provided by mixed quantum--classical (MQC) dynamics, which has found widespread application to molecular systems \cite{nelson2014, subotnik2016understanding, wang2016, crespo-otero2018recent, nelson2020}. In MQC dynamics, vibrational modes are described classically, reserving a quantum treatment for the electronic coordinates, while the quantum--classical interaction is treated self-consistently. As such, MQC dynamics provides a non-Markovian and non-perturbative treatment of the carrier--vibrational coupling, at the expense of the classical approximation taken for the vibrational modes. Notably, this approximation is exact in the short-time limit \cite{kay2006herman}, as a result of which MQC dynamics is particularly accurate for describing early phenomena, such as the scattering processes governing optical lineshapes.

In recent years, MQC dynamics (at times referred to as nonadiabatic molecular dynamics) has increasingly found applications to materials \cite{smith2019modeling}, including monolayer \cite{jiang2021real} and few-layer \cite{nie2014ultrafast, nie2015ultrafast} TMDs, as well as TMD-based heterostructures \cite{long2016quantum, zheng2018phonon, shi2020iodine}. Importantly, MQC dynamics is commonly formulated within a physical (i.e., local) basis for the involved electronic and vibrational coordinates. This is adequate for most molecular systems, but it poses a challenge for resolving the band-like excitons and phonons in TMDs, which require exceedingly-large material sizes to be included. In spite of the comparatively low cost of MQC dynamics, this significantly complicates its ability to reach convergence.

We recently proposed a reformulation of MQC dynamics within reciprocal space \cite{krotz2021reciprocal, krotz2022reciprocal}. By associating the electronic and vibrational coordinates with BZ locations, this reformulation allows MQC dynamics to be applied to band-like phenomena in solids at radically-reduced cost by truncating the BZ to only the regions providing significant contributions. BZ truncations were previously employed for converging calculations of excitons \cite{qiu2016screening} and trions \cite{tempelaar2019many-body} in TMDs. Our demonstration  of reciprocal-space MQC dynamics was restricted to  simple, one-dimensional lattices with Holstein and Peierls-type coupling of a single carrier to a single phonon branch \cite{krotz2021reciprocal, krotz2022reciprocal}. Reciprocal-space MQC dynamics has since been combined with density-functional theory and density-functional perturbation theory \cite{xie2022surface}, and has been applied to study the Floquet nonadiabatic dynamics of laser-dressed solid-state materials \cite{chen2024floquet}.

Here, we apply reciprocal-space MQC dynamics to microscopically model the optical lineshapes of the TMD MoS$_2$. To this end, we introduce a formalism invoking the Bethe--Salpeter equation (BSE) \cite{salpeter1951relativistic} with input from a parametrized band structure and a static model dielectric function, although a direct interfacing of our formalism with \textit{ab initio} calculations is possible in principle. Self-consistent coupling to a single acoustic phonon branch is incorporated by means of Ehrenfest's theorem \cite{ehrenfest1927bemerkung}, yielding a mean-field MQC framework. In order to explore convergence of our results, we introduce a truncation radius around the $K$ and $K'$ high-symmetry points within the electron--hole basis, while additionally invoking a truncation radius around the $\Gamma$ point for the total wavevector of the excitons. Combined with the BZ sampling resolution, this equips us with three convergence parameters. A fitting of calculated results for varying parameters allows us to determine asymptotic values of optical linewidths, which are found to be in good agreement with experimental measurements across a range of temperatures. We additionally find our approach to account for a sideband due to the acoustic phonon, as a result of non-Markovian effects captured by MQC dynamics. These results help establishing MQC modeling as a viable route towards studying carrier--phonon scattering in solids.

This Paper is organized as follows. In Sec.~\ref{sec:reciprocal} we briefly review the principles of reciprocal-space MQC dynamics. In Secs.~\ref{sec:absorption}, \ref{sec:excitons}, and \ref{sec:carrier-phonon} we present the theory of absorption, excitons, and carrier--phonon interactions, respectively. We then proceed with an outline of the model used to describe MoS$_2$ in Sec.~\ref{sec:model}, followed by an overview of the BZ truncations in Sec.~\ref{sec:truncations}. We then present and discuss our results in Sec.~\ref{sec:results}, after which we conclude and offer an outlook in Sec.~\ref{sec:conclusions}.

\section{Theory}\label{sec:theory}

\subsection{Reciprocal-space mixed quantum--classical dynamics}\label{sec:reciprocal}

Within MQC dynamics, classical phonons (or vibrational modes) are commonly described in terms of canonical position and momentum coordinates, which in vector form are denoted as $\bm{q}$ and $\bm{p}$, respectively. The key principle behind reciprocal-space MQC is the combining of these canonical coordinates into complex-valued coordinates, $\bm{z}$, such that $\bm{q}$ and $\bm{p}$ contribute to the real and imaginary parts, respectively \cite{krotz2021reciprocal, krotz2022reciprocal}. The complex-valued coordinates $\bm{z}$ can then be subjected to a complex Fourier transform, allowing one to rotate from a physical basis to reciprocal space \cite{krotz2021reciprocal, krotz2022reciprocal}. Upon this rotation, canonical coordinates can be recovered by deconstruction of $\bm{z}$ into canonical coordinates. While this deconstruction can be elucidating at times, it is not strictly necessary, as MQC dynamics can be fully formulated in terms of $\bm{z}$ \cite{miyazaki2024unitary}. Accordingly, the total Hamiltonian operator governing interacting carriers and phonons is partitioned as
\begin{align}
    \hat{H}(\bm{z}) = \hat{H}_{\mathrm{el}} + \hat{H}_{\mathrm{el-ph}}(\bm{z}) + H_{\mathrm{ph}}(\bm{z}).\label{eq:H_tot}
\end{align}
Here, $\hat{H}_{\mathrm{el}}$ is the Hamiltonian operator describing the electronic subsystem and $H_{\mathrm{ph}}(\bm{z})$ is the Hamiltonian function of the phonons. The interaction between electronic carriers and phonons is described by the operator $\hat{H}_{\mathrm{el-ph}}(\bm{z})$, which depends parametrically on the phonon coordinates.

The phonon coordinates are propagated by means of the Hamilton equations of motion, which for the complex variables take the form \cite{miyazaki2024unitary}
\begin{align}
    \dot{\bm{z}} &= -i \; \nabla_{\bm{z}^*} \braket{\hat{H}} \nonumber \\
    &= -i \; \nabla_{\bm{z}^*} \left(\braket{\hat{H}_{\mathrm{el-ph}}} + H_{\mathrm{ph}}\right).
\end{align}
Here, $\braket{\ldots}$ denotes the expectation value through which the quantum subsystem provides a contribution to the classical dynamics. Within mean-field MQC dynamics, this expectation is taken with respect to the electronic wavefunction, $\Psi$, as predicted by Ehrenfest's theorem \cite{ehrenfest1927bemerkung}. Accordingly, we have for the electron--phonon Hamiltonian operator
\begin{align}
    \braket{\hat{H}_{\mathrm{el-ph}}} = \braket{\Psi|\hat{H}_{\mathrm{el-ph}}|\Psi}.
\end{align}
We note that such a mean-field approximation is known to suffer from over-thermalization \cite{parandekar2005mixed, parandekar2006detailed}, which at long times leads to inaccurate dynamics as well as difficulties in applying BZ truncations \cite{krotz2022reciprocal}. However, the absorption processes considered in the present work ensue well before thermalization sets in, so that these issues are expected to be minor. The electronic wavefunction, on the other hand, is propagated through the time-dependent Schr\"odinger equation,
\begin{align}
    i\hbar\ket{\dot{\Psi}} &= \hat{H}\ket{\Psi} \nonumber \\
    &= \left( \hat{H}_{\mathrm{el}} + \hat{H}_{\mathrm{el-ph}} \right) \ket{\Psi}.
\end{align}

Typically, the electronic wavefunction is initialized in some given state, while the phonon coordinates are sampled from a thermal distribution, after which the phononic and electronic equations of motion can be self-consistently solved. Properties of interest are then calculated as an average over quantum--classical trajectories.

\subsection{Absorption}\label{sec:absorption}

Given a semiconducting material, the (electronic) absorption spectrum for a given polarization direction $\lambda$ follows from a Fourier transform of the response function, $R_{\lambda}(t)$, as
\begin{align}
    A_{\lambda}(\omega) = \int_{0}^{\infty} \dd \tau\, e^{i\omega \tau}R_{\lambda}(\tau),
    \label{eq:abs}
\end{align}
with the response function given by
\begin{align}
    R_{\lambda}(\tau) = \braket{0 | \hat{P}_{\lambda} | \Psi(\tau)}. \label{eq:resp_func}
\end{align}
Here, $\ket{0}$ represents the electronic ground state, with completely-filled valence bands and empty conduction bands (referred to as the Fermi vacuum), and $\Psi(\tau)$ represents the electronic wavefunction at time $\tau$.

The electronic momentum operator appearing in Eq.~\ref{eq:resp_func} is given by \cite{berkelbach2015bright, salij2021microscopic}
\begin{align}
    \hat{P}_\lambda = \frac{m_{\mathrm{e}}}{\hbar} \sum_{\bm k,v,c} \braket{ \psi_{\bm k,v} | \, \bm{e}_\lambda \cdot \nabla_{\bm k} \hat{h}_{\bm k} \, | \psi_{\bm k,c} } \hat{c}^{\dagger}_{\bm k,v} \hat{c}_{\bm k,c},
    \label{eq:momentum_operator}
\end{align}
where $m_{\mathrm{e}}$ is the electron rest mass and $\bm{e}_\lambda$ is the unit vector along direction $\lambda$. The operator $\hat{h}_{\bm k}$ represents the noninteracting quasiparticle Hamiltonian. Its eigensolutions,
\begin{align}
    \hat{h}_{\bm k} \ket{\psi_{\bm k,c(v)}} = \epsilon_{\bm k,c(v)} \ket{\psi_{\bm k,c(v)}},
    \label{eq:h_k}
\end{align}
define the quasiparticle spin-bands, where $\bm k$ denotes the wavevector, and where $c(v)$ represents the conduction (valence) spin-band label. In Eq.~\ref{eq:momentum_operator}, $\hat{c}^{\dagger}_{\bm k,c(v)}$ and $\hat{c}_{\bm k,c(v)}$ represent the operators for creation and annihilation, respectively, of an electron in spin-band $c(v)$ with wavevector $\bm k$.

At time $\tau = 0$, the electronic wavefunction is initialized as
\begin{align}
    \ket{\Psi(\tau=0)} = \hat{P}_{\lambda}^{\dagger}\ket{0}.\label{eq:psi_0}
\end{align}
In governing its subsequent evolution, we resort to reciprocal-space MQC dynamics, as discussed in the Sec.~\ref{sec:reciprocal}, under application of the Hamiltonians presented in Secs.~\ref{sec:excitons} and \ref{sec:carrier-phonon}.

\subsection{Excitons}\label{sec:excitons}

Excitonic states in a semiconducting material are governed by the BSE, which can be expressed in Hamiltonian form as
\begin{align}
    \hat{H}_{\mathrm{el}} = & \sum_{\bm{k}_{1},\bm{k}_{2}} \sum_{c,v} (\epsilon_{\bm{k}_{1},c} - \epsilon_{\bm{k}_{2},v}) \hat{c}_{\bm{k}_{1},c}^\dagger \hat{c}_{\bm{k}_{2},v} \hat{c}_{\bm{k}_{2},v}^\dagger \hat{c}_{\bm{k}_{1},c} \label{eq:H_el_1} \\
    &+ \frac{1}{A} \sum_{\bm{k}_{1},\bm{k}_{2},\bm{\kappa}}\sum_{c,v,c',v'} \braket{\psi^{\dagger}_{\bm{k}_{2}+\bm{\kappa},v'}\psi_{\bm{k}_{1}+\bm{\kappa},c'}| \hat{K}_{\mathrm{int}}| \psi^{\dagger}_{\bm{k}_{2},v}\psi_{\bm{k}_{1},c}} \nonumber \\
    &\times \hat{c}^{\dagger}_{\bm{k}_{1}+\bm{\kappa},c'}\hat{c}_{\bm{k}_{2}+\bm{\kappa},v'}\hat{c}_{\bm{k}_{2},v}^{\dagger}\hat{c}_{\bm{k}_{1},c}.\nonumber
\end{align}
Here, $\hat{K}_{\mathrm{int}}$ represents the electron--hole interaction operator, and $A$ is a normalization to the length, area, or volume of the involved material. The daggers appearing in the interaction kernel refer to hole states, being Hermitian conjugates of electron states.

Excitonic eigenstates of the electronic Hamiltonian operator, satisfying $\hat{H}_{\mathrm{el}} \ket{\Phi_n} = E_n \ket{\Phi_n}$, can be expanded in the basis of electron and hole excitations of the Fermi vacuum as
\begin{align}
    \ket{\Phi_n} = \sum_{\bm{k},c,v} A_{\bm{k},c,v}^n \hat{c}^{\dagger}_{\bm{k} + \bar{\bm{k}}_n,c} \hat{c}_{\bm{k},v}\ket{0},
\end{align}
where the expansion coefficients are denoted $A_{\bm{k},c,v}^n$, and where $\bar{\bm{k}}_n$ is the total wavevector of exciton $n$.

\subsection{Carrier--phonon interactions}\label{sec:carrier-phonon}

Under the harmonic approximation, a general form of the Hamiltonian function governing phonons in materials is given by \cite{krotz2021reciprocal}
\begin{align}
    H_{\mathrm{ph}} = \sum_{\bm{k},\mu} \hbar \omega_{\bm{k},\mu} z_{\bm{k},\mu}^* z_{\bm{k},\mu},\nonumber
\end{align}
where $\bm{k}$ and $\mu$ denote the phonon wavevector and branch, respectively, whereas $\omega_{\bm{k},\mu}$ represents the associated harmonic frequency. In adopting a representation of classical phonons in terms of complex coordinates (cf.~Sec.~\ref{sec:reciprocal}), we have taken $z_{\bm{k},\mu}$ to correspond to the eigenvalue associated with the coherent state of the harmonic oscillator \cite{kim2022coherent} representing the phonon with wavevector $\bm{k}$ and branch $\mu$. This coordinate can be decomposed as
\begin{align}
    z_{\bm{k},\mu} \equiv \sqrt{\frac{\omega_{\bm{k},\mu}}{2\hbar}}\left( q_{\bm{k},\mu} + i \frac{p_{\bm{k},\mu}}{\omega_{\bm{k},\mu}} \right),
\end{align}
through which relevant equations can be expressed in terms of canonical coordinates $q_{\bm{k},\mu}$ and $p_{\bm{k},\mu}$, as mentioned in Sec.~\ref{sec:reciprocal}. In the following we will refrain from doing so, as relevant equations take a more intuitive form when expressed in terms of $z_{\bm{k},\mu}$.

The carrier--phonon interactions are governed by the Hamiltonian
\begin{align}
    \hat{H}_{\mathrm{el-ph}} = \sum_{\bm{k}_{1},\bm{k}_{2},\bm{\kappa}} &\, \sum_{c,v,\mu} \Big(g^{c}_{\bm{k}_{1},\bm{\kappa},\mu} \hat{c}^{\dagger}_{\bm{k}_{1}+\bm{\kappa},c}\hat{c}_{\bm{k}_{2},v}\hat{c}^{\dagger}_{\bm{k}_{2},v}\hat{c}_{\bm{k}_{1},c}\label{eq:H_elph_1} \nonumber \\
    - &\, g_{\bm{k}_{2},\bm{\kappa},\mu}^{v} \hat{c}^{\dagger}_{\bm{k}_{1},c}\hat{c}_{\bm{k}_{2},v}\hat{c}^{\dagger}_{\bm{k}_{2}+\bm{\kappa},v}\hat{c}_{\bm{k}_{1},c} \Big)
    \left(z^{*}_{-\bm{\kappa},\mu} +z_{\bm{\kappa},\mu}\right).
\end{align}
Here, $g^{c(v)}_{\bm{k},\bm{\kappa},\mu}$ is the spin-conserving intraband matrix element for an electron (hole) scattering onto a phonon in branch $\mu$ with wavevector $\bm{\kappa}$, thereby undergoing a change of wavevector from $\bm{k}$ to $\bm{k}+\bm{\kappa}$.\footnote{We assume a sign convention where $g^{c}_{\bm{k},\bm{\kappa},\mu}$ and $g^{v}_{\bm{k},\bm{\kappa},\mu}$ have the same sign if their coupling mechanisms depend on the charge of the carrier, and opposite sign if the mechanism is charge-independent.} As elaborated upon in Refs.~\citenum{selig2016ExcitonicLinewidthCoherence} and \citenum{selig2018exciton}, interband scattering as well as spin-nonconserving carrier--phonon scattering processes provide negligible contributions to optical linewidths due to the large energetic separation between bands and the comparatively-long timescales associated with spin flips. These scattering processes are therefore omitted here.

Restricting ourselves to the Hilbert space spanned by single electron--hole pairs, the total Hamiltonian can be transformed into the excitonic eigenbasis. The advantage of this is that the electronic Hamiltonian can be solved for \textit{a priori}, following which the total Hamiltonian can be diagonalized within a reduced set of excitonic eigenstates (see Sec.~\ref{sec:truncations}). Accordingly, the exciton creation operator is first defined as $\hat{C}^{\dagger}_{n}\ket{0} \equiv \ket{\Phi_{n}}$, and analogously for the exciton annihilation operator. In terms of these operators, the electronic Hamiltonian is given by
\begin{align}
    \hat{H}_{\mathrm{el}} = \sum_n E_n \hat{C}^{\dagger}_n\hat{C}_n.
\end{align}
Inserting the equality $\hat{c}^{\dagger}_{\bm{k}+\bar{\bm{k}},c}\hat{c}_{\bm{k},v}=\sum_{n}A^{n*}_{\bm{k},c,v}\hat{C}^{\dagger}_{n}$ into Eq.~\ref{eq:H_elph_1} yields for the carrier--phonon interaction Hamiltonian
\begin{align}
    \hat{H}_{\mathrm{el-ph}} = \sum_{n,m,\mu}G_{n,m,\mu} \hat{C}^{\dagger}_n \hat{C}^{\phantom{\dagger}}_m \left(z^{*}_{-\bar{\bm{k}}_{nm},\mu} + z_{\bar{\bm{k}}_{nm},\mu}\right),
    \label{eq:ex_ph}
\end{align}
where $\bar{\bm{k}}_{nm} \equiv \bar{\bm{k}}_{n}-\bar{\bm{k}}_{m}$ is the difference in total wavevectors of excitons $n$ and $m$ and where
\begin{align}
    G_{n,m,\mu} &\equiv \sum_{\bm{k},c,v} \Big[g_{\bm{k}+\bar{\bm{k}}_{m},\bar{\bm{k}}_{nm},\mu}^{c} \left(A^n_{\bm{k},c,v}\right)^* A^m_{\bm{k},c,v} \label{eq:Ex-ph} \\
    &- g^{v}_{\bm{k}-\bar{\bm{k}}_{n},\bar{\bm{k}}_{nm},\mu} \left(A^n_{\bm{k}-\bar{\bm{k}}_{n},c,v}\right)^* A^m_{\bm{k}-\bar{\bm{k}}_{m},c,v} \Big]\nonumber
\end{align}
is the exciton--phonon interaction element.

\subsection{Model for MoS\texorpdfstring{\textsubscript{2}}{2}}\label{sec:model}

\begin{table}[b!]
\centering
\begin{tabular}{ |c|c|c| } 
 \hline
 Parameter &  Value & Ref.\\ 
 \hline 
 $a$ & 3.19 \AA & ~\citenum{xiao2012coupled}~ \\ 
 $t$ & 1.10 eV & ~\citenum{xiao2012coupled}~ \\ 
 $\Delta$ & 1.66 eV & ~\citenum{xiao2012coupled}~ \\ 
 $\lambda_{\mathrm{v}}$ & $148$ meV & ~\citenum{tempelaar2019many-body}~ \\ 
 $\lambda_{\mathrm{c}}$ & $-3.0$ meV & ~\citenum{tempelaar2019many-body}~ \\ 
 $\chi_{\mathrm{2D}}$ & $6.60$ \AA & ~\citenum{tempelaar2019many-body}~ \\ 
 $v_{\mathrm{LA}}$ & $66$ \AA~ps$^{-1}$ & ~\citenum{li2013IntrinsicElectricalTransporta}~ \\ 
 $m$ & $2.66 \cdot 10^{-25}$ kg & -- \\
 $D_{1}^{c}$ & $4.5$ eV & ~\citenum{li2013IntrinsicElectricalTransporta}~ \\ 
 $D_{1}^{v}$ & $2.5$ eV & ~\citenum{li2013IntrinsicElectricalTransporta}~ \\
 \hline
\end{tabular}
\caption{Parametrization applied in our calculations, including the references where parameters were taken from. The unit cell mass $m$ was taken to be the total mass of one molybdenum atom and two sulphur atoms.}
\label{tab:parms}
\end{table}

The combination of Secs.~\ref{sec:reciprocal} -- \ref{sec:carrier-phonon} presents a comprehensive and microscopic formalism that can in principle be interfaced with an \textit{ab initio} treatment of a material including interacting carriers and phonons. While we consider such interfacing of interest for future research, here we instead resort to a model band structure as well as models for the electron--hole and carrier-phonon interaction elements, each parametrized against \textit{ab initio} calculations, in order to simplify the implementation of our approach. All applied parameters are shown in Tab.~\ref{tab:parms}.

Computational studies \cite{xiao2012coupled, berkelbach2015bright} have shown the conduction and valence bands of MoS$_2$ in the vicinity of the bandgap regions to be well-represented by an appropriately-parametrized massive Dirac-like Hamiltonian of the form \cite{xiao2012coupled, ochoa2013spin}
\begin{align}
    \hat{h}_{\bm{k}} = \left(\begin{array}{cc} -\Delta/2+\lambda_\text{v}\tau_{\bm{k}}\hat{S}_z & at(\tau_{\bm{k}}\tilde{k}_x+i\tilde{k}_y) \\ at(\tau_{\bm k}\tilde{k}_x-i\tilde{k}_y) & \Delta/2+\lambda_\text{c}\tau_{\bm k}\hat{S}_z \end{array}\right).\label{eq:H_k}
\end{align}
Here, $\Delta$ is the band gap, $a$ is the lattice constant, and $t$ is the effective nearest-neighbor transfer integral.
Furthermore, $\tilde{\bm{k}}$ denotes the wavevector \textit{relative} to the $K$ or $K'$ high-symmetry point, whichever is closest, and $\tau_{\bm{\kappa}} = +1$ ($-1$) if $\bm{k}$ is closer to $K$ ($K'$). 
Spin-orbit splitting of the conduction (valence) band by the amount of $2\lambda_{c(v)}$ is incorporated, with $\hat{S}_{z}$ being the out-of-plane spin operator.
Solving Eq.~\ref{eq:h_k} with substitution of Eq.~\ref{eq:H_k} provides the quasiparticle energies $\epsilon_{\bm{k},c(v)}$ and wavefunctions $\psi_{\bm{k},c(v)}$ necessary to construct the BSE Hamiltonian, Eq.~\ref{eq:H_el_1}.

The electron--hole interaction elements appearing in the BSE Hamiltonian include contributions from a screened direct term and an unscreened exchange term. As with effective mass models \cite{berkelbach2013theory}, parametrized band structures such as the one applied here lack a detailed structure of the atomic orbitals. As is commonly done \cite{zhang2014absorption, konabe2014effect, berkelbach2015bright, tempelaar2019many-body}, we therefore approximate the interaction elements as 
\begin{align}
    & \braket{\psi^{\dagger}_{\bm{k}_{2}+\bm{\kappa},v'}\psi_{\bm{k}_{1}+\bm{\kappa},c'}| \hat{K}_{\mathrm{int}}| \psi^{\dagger}_{\bm{k}_{2},v}\psi_{\bm{k}_{1},c}} \\
    & =-\braket{\psi_{\bm{k}_{1}+\bm{\kappa},c'} |   \psi_{\bm{k}_{1},c}} \braket{\psi_{\bm{k}_{2},v} | \psi_{\bm{k}_{2}+\bm{\kappa},v'}} W(\bm{\kappa}),\nonumber \\
    & + \braket{\psi_{\bm{k}_{1}+\bm{\kappa},c'}|\psi_{\bm{k}_{2}+\bm{\kappa},v'}}\braket{\psi_{\bm{k}_{2},v}|\psi_{\bm{k}_{1},c}}v(\bm{k}_{1}-\bm{k}_{2}),\nonumber
\end{align}
which also neglects the frequency-dependence of the material's inverse dielectric function. Here, $W(\bm{k})$ is the screened Coulomb interaction, which is modeled by means of the Rytova--Keldysh potential \cite{rytova1967screened, keldysh1979coulomb},
\begin{align}
    W(\bm{k}) = \frac{e^2}{2\epsilon_0 \vert\bm{k}\vert(1+2\pi\chi_\text{2D}\vert\bm{k}\vert)}.
\end{align}
Here, $e$ is the elementary charge, $\epsilon_{0}$ is the vacuum permittivity, and $\chi_{\mathrm{2D}}$ is the two-dimensional polarizability. The unscreened Coulomb interaction in two dimensions, $v(\bm{k})$, is recovered from the Rytova--Keldysh potential in the limit of vanishing $\chi_{\mathrm{2D}}$ as $v(\bm{k}) = e^{2}/(2\epsilon_{0}\vert\bm{k}\vert)$. 

The BSE Hamiltonian based on the two-band model and the aforementioned model interactions was previously shown to produce results in good agreement with measurements. For MoS$_2$, it was shown to yield trion binding energies \cite{tempelaar2019many-body} consistent with experimental observations \cite{mak2013tightly, zhang2015on} and to reproduce \cite{berkelbach2015bright} the experimentally-observed non-hydrogenic excitonic Rydberg series \cite{chernikovExcitonBindingEnergy2014}, while showing good overall agreement with a more sophisticated three-band model \cite{liu2013three}. For the related monolayer TMD material WS$_2$, it was shown to reproduce \cite{berkelbach2015bright} the experimentally-observed two-photon absorption spectrum \cite{ye2014probing}, while it was also found to capture \cite{tempelaar2019many-body} the coherent exciton--trion signals measured in two-dimensional spectroscopy of MoSe$_2$ \cite{hao2016coherent}.

In Ref.~\citenum{selig2016ExcitonicLinewidthCoherence}, Knorr and coworkers have shown optical phonons to minimally affect the temperature-dependent component of the optical linewidths of MoSe$_2$, with the predominant contribution being provided by acoustic phonons instead. Although MoS$_2$ was not evaluated in this work, it was argued that a similar linewidth behavior is to be expected based on structural similarities with MoSe$_2$ \cite{selig2016ExcitonicLinewidthCoherence}. In simplifying our model, we therefore restrict ourselves to the acoustic phonon contributions. While Ref.~\citenum{selig2016ExcitonicLinewidthCoherence} described the longitudinal acoustic (LA) and transverse acoustic (TA) modes as two degenerate phonon branches, we found that numerically-identical dynamics are obtained when combining both modes into a single branch with a deformation potential constant multiplied by $\sqrt{2}$. We have therefore adopted a single-branch model with frequency $\omega_{\bm{\kappa}}= v_{\mathrm{LA}}\vert \bm{\kappa}\vert$, where $v_{\mathrm{LA}}$ is the sound velocity of the LA mode \cite{selig2016ExcitonicLinewidthCoherence}. Here and henceforth, the branch index is omitted in the equations, being redundant in this single-branch scenario. \textit{Ab initio} calculated values of the deformation potential constants contain both deformation potential and piezoelectric couplings, which are approximately equal in contribution \cite{selig2016ExcitonicLinewidthCoherence}. However, the piezoelectric contribution will experience a destructive interference in the net exciton--phonon interaction element as piezoelectric couplings involving electrons and holes have opposite signs due to their dependence on carrier charge. Similarly to Ref.~\citenum{lengersTheoryAbsorptionLine2020}, we therefore omit the piezoelectric contribution by dividing the \textit{ab initio} calculated deformation potential constant by $\sqrt{2}$, which cancels against the aforementioned $\sqrt{2}$ multiplication.

A generic expression of the carrier--phonon interaction elements is given by \cite{kaasbjergPhononlimitedMobilityType2012}
\begin{align}
    g^{c(v)}_{\bm{k},\bm{\kappa}} =  \sqrt{\frac{\hbar}{2mN\omega_{\bm{\kappa}}}} \braket{\psi_{\bm{k}+\bm{\kappa}, c(v)}|\Delta_{\bm{k}}\hat{V}|\psi_{\bm{k}, c(v)}},
    \label{eq:G_eph}
\end{align}
where $m$ is the mass of the unit cell, $N$ is the total number of unit cells, and $\Delta_{\bm{k}}\hat{V}$ is the change in the effective potential per unit displacement along the phonon coordinate. In Ref.~\citenum{selig2016ExcitonicLinewidthCoherence}, the couplings between carriers and acoustic phonons were described by means of a first-order deformation potential approach, which amounts to the replacement
\begin{align}
    \braket{\psi_{\bm{k}+\bm{\kappa}, c(v)}|\Delta_{\bm{k}}\hat{V}|\psi_{\bm{k}, c(v)}} \leftarrow D_1^{c(v)}\vert\bm{\kappa}\vert.
    \label{eq:def_pot}
\end{align}
Here, $D_1^{c(v)}$ is the deformation potential constant, which is treated as a parameter. We note that the left-hand side of Eq.~\ref{eq:def_pot} is subject to a global gauge freedom introduced by the quasiparticle eigenstates, which is eliminated in the replacement. In order to reinstate this gauge, as necessary for integration within the BSE, we instead adopt a modified deformation potential approach, by applying the replacement
\begin{align}
    \braket{\psi_{\bm{k}+\bm{\kappa}, c(v)}|\Delta_{\bm{k}}\hat{V}|\psi_{\bm{k}, c(v)}} \leftarrow D_1^{c(v)}\vert\bm{\kappa}\vert \exp\left({i\phi^{c(v)}_{\bm{k},\bm{\kappa}}}\right),
\end{align}
where $\phi^{c(v)}_{\bm{k},\bm{\kappa}} \equiv \arg(\braket{\psi_{\bm{k}+\bm{\kappa},c(v)}\vert\psi_{\bm{k},c(v)}})$ accounts for the gauge.

\subsection{Brillouin zone truncations}\label{sec:truncations}

\begin{figure}[t!]
    \centering
    \includegraphics{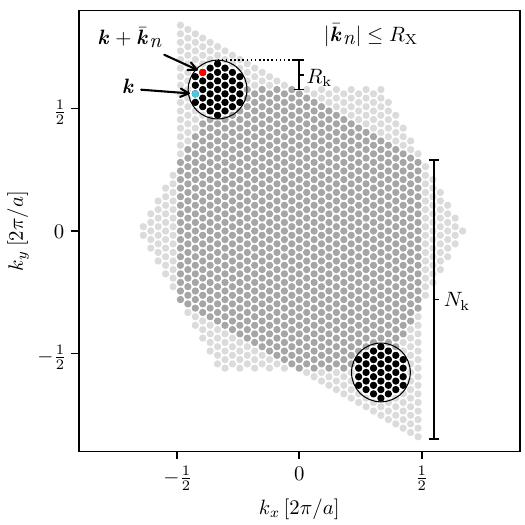}
    \caption{Depiction of the hexagonal BZ of MoS$_2$, with an equivalent Monkhorst--Pack grid overlaid. Discrete reciprocal-space lattice points are shown for a resolution of $N_{\mathrm{k}} = 33$. Also depicted is the truncation radius applied to truncate the electron--hole basis, $R_{\mathrm{k}}$, which limits the lattice points to within a circle (shown in black) centered at the $K$ and $K'$ points (the corners of the hexagonal BZ). As a further means of truncation, the total wavevector of an exciton is limited as $\vert \bar{\bm{k}}_n \vert < R_{\mathrm{X}}$. This wavevector is the difference of that of the electron $\bar{\bm{k}}_n + \bm{k}$ (shown as red) and that of the hole $\bm{k}$ (shown as cyan).}
    \label{fig:FIG_conv_params}
\end{figure}

Critical to our efforts to push our modeling towards convergence is our ability to perform efficient truncations of the BZ within the applied formalism. Within the realm of MQC dynamics, such truncations are uniquely enabled by our reciprocal-space formalism \cite{krotz2021reciprocal, krotz2022reciprocal}. Fig.~\ref{fig:FIG_conv_params} depicts various BZ truncations taken. As shown here, the hexagonal BZ of MoS$_2$ is described by means of a Monkhorst--Pack grid, using a resolution of $N_{\mathrm{k}}\times N_{\mathrm{k}}$. Without any truncations, the number of excitonic basis states scales as $N_{\mathrm{k}}^4$.

As a first means of performing truncations, the grid points included in the electron--hole basis used to solve the BSE Hamiltonian are restricted to those within a given radius around the $K$ and $K'$ points, denoted $R_{\mathrm{k}}$. This truncation relies on the key principle that optical absorption produces carriers close to those high-symmetry points \cite{berkelbach2015bright}, and was previously applied in static calculations of excitons \cite{qiu2016screening} and trions \cite{tempelaar2019many-body} in MoS$_2$.

Solving for the BSE Hamiltonian yields a manifold of discrete exciton bands at low energies, in addition to a continuum of high energy states \cite{wu2015ExcitonBandStructurea}. Restricting ourselves to excitons where the constituent electron and hole carry the same spin, we find the four bands at lowest energy, shown in Fig.~\ref{fig:scheme} (c), to correspond to the spin-up and spin-down electron--hole pair located in the $K$ and $K'$ valleys, respectively. At the $\Gamma$ point with $\bar{\bm{k}}_n = 0$, these bands form degenerate pairs, and represent the lowest state within the nonhydrogenic Rydberg series (with the other states in this series residing at higher energies). With increasing $\vert\bar{\bm{k}}_n\vert$, the degenerate pairs split, as electron--hole exchange interactions lift one of the bands above the other \cite{wu2015ExcitonBandStructurea}. In our model we include only the lowest band, which is unaffected by exchange interactions \cite{rohlfingElectronholeExcitationsOptical2000}.

Additionally, a truncation radius is invoked around the $\Gamma$ point for the total wavevector of the excitons, denoted $R_{\mathrm{X}}$, such that $\vert \bar{\bm{k}}_n \vert < R_{\mathrm{X}}$. The latter has ramifications for the phononic coordinates, since phonons are associated with changes in excitonic wavevector, as per Eq.~\ref{eq:ex_ph}. In our calculations, only coordinates are included that contribute to excitonic scattering processes allowed by our truncated electron--hole basis, which implies a truncation of phonon wavevectors to within a radius around the $\Gamma$ point.

Altogether, we find ourselves with three parameters, $N_{\mathrm{k}}$, $R_{\mathrm{k}}$, and $R_{\mathrm{X}}$, with respect to which convergence can be feasibly modulated.

\section{Results and Discussion}\label{sec:results}

The optical linewidth of MoS$_2$ receives both temperature-dependent and temperature-independent contributions. In the following, we will only consider the temperature-dependent contribution, which is attributed to acoustic phonons \cite{selig2016ExcitonicLinewidthCoherence}.

\begin{figure}[b!]
    \centering
    \includegraphics{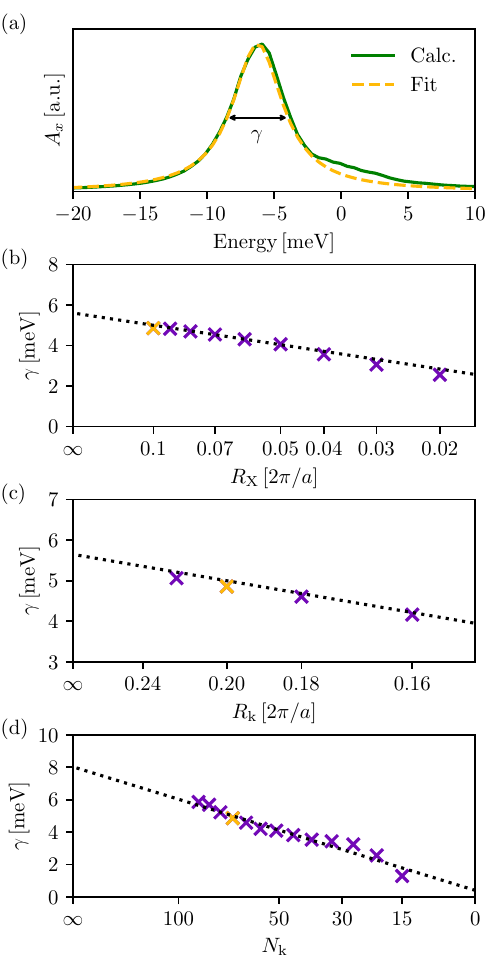}
    \caption{(a) Low-energy absorption spectrum of MoS$_2$ at 300 K, calculated with truncation parameters $N_{\mathrm{k}} = 69$, $R_{\mathrm{k}} = 0.2$, and $R_{\mathrm{X}} = 0.1$ (green, solid). Also shown is a Lorentzian fit to the spectral profile below the maximum of the principal peak (orange, dashed), emphasizing the phonon sideband at higher energy. The associated FWHM is used to determine the linewidth $\gamma$, as indicated. Energy is normalized to that of the lowest exciton at the $\Gamma$ point (cf.~Fig.~\ref{fig:scheme}).
    (b) Calculated linewidths at varying $R_{\mathrm{X}}$ and with $N_{\mathrm{k}} = 69$ and $R_{\mathrm{k}} = 0.2$ (purple markers), shown together with a fit (dotted, see text for details). Horizontal axis is scaled according to the fit function. (c) As in (b) but for varying $R_{\mathrm{k}}$. (d) As in (b) but for varying $N_{\mathrm{k}}$. Identical data across panels is shown in orange.}
    \label{fig:FIG_conv_slices}
\end{figure}

Shown in Fig.~\ref{fig:FIG_conv_slices} (a) is the $x$ polarized absorption spectrum of MoS$_2$ at a temperature of $T = 300~\mathrm{K}$ calculated using $N_{\mathrm{k}} = 69$, $R_{\mathrm{k}} = 0.2$, and $R_{\mathrm{X}} = 0.1$. (Here and henceforth, distances in reciprocal-space are given in units of $2\pi /a$. Moreover, the $x$ polarization was chosen arbitrarily, as any linear polarization direction will yield formally-identical results.) To calculate this spectrum, the electronic wavefunction was initialized according to Eq.~\ref{eq:psi_0} (with $\lambda = x$), while the classical phonon coordinates were stochastically drawn from a thermal Boltzmann distribution. Within the complex parameter representation, this amounts to drawing the magnitude of the parameters according to the probability
\begin{align}
    P\left(\vert z_{\bm{k}} \vert\right) \propto \exp\left( - \beta \hbar \omega_{\bm{k}} \vert z_{\bm{k}} \vert ^2\right),
\end{align}
with $\beta = 1 / k_{\mathrm{B}} T$ as the inverse temperature, while the argument of the variable is uniformly drawn from the interval $[0,2\pi)$. Upon this initialization, dynamics was computed over the course of $7.82$ picoseconds with a timestep of $0.26$ femtoseconds for a total of $20,000$ trajectories, yielding the response function according to Eq.~\ref{eq:resp_func}, which was then Fourier transformed in order to produce the absorption spectrum as per Eq.~\ref{eq:abs}.

The resulting spectrum shown in Fig.~\ref{fig:FIG_conv_params} (a) consists of one principal peak accompanied by a broad sideband at a blueshift of roughly 7 meV. The emergence of such sideband is a manifestation of non-Markovian dynamics \cite{christiansen2017phonon, lengersTheoryAbsorptionLine2020}, as accounted for by our MQC approach, and arises due to coupling of the exciton to the acoustic phonon branch. We note that the intensity and blueshift are small compared to previously-reported phonon sidebands, which were attributed to optical phonons \cite{christiansen2017phonon} not included in our model. As such, we expect the comparatively-minor sideband due to the acoustic phonon branch predicted by our modeling to be camouflaged by those from the optical phonons, while the acoustic phonon contribution instead affects the linewidths \cite{selig2016ExcitonicLinewidthCoherence}. In order to determine the linewidth of the principal peak, without interference from the phonon sideband, we performed a Lorentzian fit to the low-energy half of this peak, as shown in Fig.~\ref{fig:FIG_conv_params} (a). The full width at half maximum (FWHM) of the Lorentzian function is then used to determine the linewidth, which in Fig.~\ref{fig:FIG_conv_params} (a) amounts to roughly 5~meV. (We have also performed a fitting to the entire lineshape, including sideband, which yields very similar results.)
    
To explore linewidth values in the asymptotic limits of $N_{\mathrm{k}}$, $R_{\mathrm{k}}$, and $R_{\mathrm{X}}$, we performed calculations at varying parameters, and fitted to the obtained linewidths the function
\begin{align}
    \tilde{\gamma}(N_{\mathrm{k}}, R_{\mathrm{k}}, R_{\mathrm{X}}) = \tilde{\gamma}_{\infty}f_{\mathrm{N}}(N_{\mathrm{k}})f_{\mathrm{k}}(R_{\mathrm{k}})f_{\mathrm{X}}(R_{\mathrm{X}}),
    \label{eq:fit}
\end{align}
where $f_{\mathrm{N}}(N_{\mathrm{k}}) \equiv 1 - \exp(-b_{\mathrm{N}} N_{\mathrm{k}} + c_{\mathrm{N}})$, and where $f_{\mathrm{k}}(R_{\mathrm{k}})$ and $f_{\mathrm{X}}(R_{\mathrm{X}})$ are defined similarly. This function relies on the assumption that the convergence behaviors associated with the parameters is uncorrelated.
    
Shown in Fig.~\ref{fig:FIG_conv_slices} (b), (c), and (d) are slices of the calculated linewidth data as a function of $R_{\mathrm{X}}$, $R_{\mathrm{k}}$, and $N_{\mathrm{k}}$, respectively, while keeping the other two parameters fixed at the values from Fig.~\ref{fig:FIG_conv_slices} (a). Also shown is the fit based on Eq.~\ref{eq:fit}. For the $R_{\mathrm{X}}$ slice, the $x$ axis depicts $f_{\mathrm{X}}(R_{\mathrm{X}})$, such that the origin corresponds to the asymptotic limit and the exponential fit produces a straight line. A similar depiction is adopted for the $R_{\mathrm{k}}$ and $N_{\mathrm{k}}$ slices. For all three slices, the calculated data is seen to closely follow the fitted curves, indicating that convergence with respect to $R_{\mathrm{X}}$, $R_{\mathrm{k}}$, and $N_{\mathrm{k}}$ is indeed largely uncorrelated and exponential.
    
As seen in Fig.~\ref{fig:FIG_conv_slices} (b), $R_{\mathrm{X}} = 0.1$ is sufficient to approach convergence. The relatively small convergence radius necessary can be rationalized by recognizing that phonons absorb energy when scattering with the exciton, in addition to lattice momentum. The conservation of both energy and lattice momentum is most closely satisfied when the phonon and exciton dispersions are similar, which for the acoustic phonon branch occurs near the $\Gamma$ point, as seen in Fig.~\ref{fig:scheme} (c). As such, effective exciton--phonon scattering pathways couple exciton states within a relatively-small radius around the $\Gamma$ point.

Convergence is markedly slower for $R_{\mathrm{k}}$, as seen in Fig.~\ref{fig:FIG_conv_slices} (c), with radii exceeding $0.2$ being necessary to approximate the asymptotic limit. Notably, this amounts to twice the radius used previously to converge the binding energy of the optically-accessible lowest-energy exciton \cite{tempelaar2019many-body}. This difference is rationalized by appreciating that optically-accessible excitons have $\bar{\bm{k}}_n = 0$, i.e., are fully described with $R_{\mathrm{X}} = 0$. While $R_{\mathrm{k}} \sim 0.1$ suffices in this case, carrier--phonon couplings induce a scattering of the total exciton wavevector to within $R_{\mathrm{X}} = 0.1$. Since this total wavevector can be absorbed in the electron, the hole, or both, this implies an enhancement of the carrier wavevector distribution radius to $R_{\mathrm{k}} \sim 0.2$.

Reaching convergence for $N_{\mathrm{k}}$ is decidedly more demanding than for $R_{\mathrm{X}}$ or $R_{\mathrm{k}}$. At small values of $N_{\mathrm{k}}$, i.e., at course sampling of the BZ, the continuous exciton band becomes discretized into states with well-separated energies and wavevectors, which affects the dynamics. This effect is particularly pronounced at elevated temperatures, where higher regions of the (quasi)parabolic exciton dispersion are accessible, and where steeper dispersions lead to larger gaps between discrete states. For that reason, we found convergence with respect to $N_{\mathrm{k}}$ to be markedly faster at lower temperatures.
    
While each of the slices shown in Fig.~\ref{fig:FIG_conv_slices} (b-d) only provide the asymptotic limit with respect to a single parameter, it is $\tilde{\gamma}_{\infty}$ obtained through the fitting function (Eq.~\ref{eq:fit}) that represents the linewidth in the asymptotic limit for all three parameters simultaneously. While the single calculation presented in Fig.~\ref{fig:FIG_conv_slices} (a) resulted in a linewidth of $\sim$5~meV, the simultaneous fitting to all calculated data yields a linewidth of $\tilde{\gamma}_{\infty}\sim 10$~meV. This value is in near-perfect agreement with experimental measurements of the temperature-dependent contribution to the optical linewidth at 300~K \cite{dey2016OpticalCoherenceAtomicMonolayer}.

\begin{figure}
    \centering
    \includegraphics{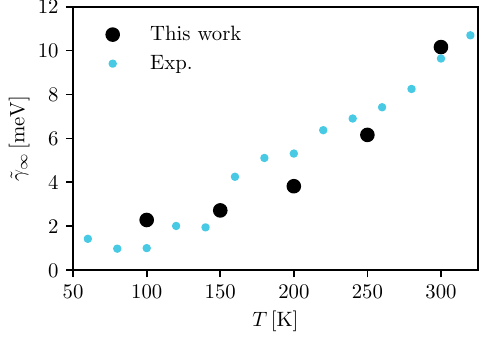}
    \caption{Asymptotic linewidth of MoS$_2$ obtained through a fitting to calculations at varying truncation parameters, as a function of temperature (black markers). Also shown are experimentally-measured linewidths (cyan) \cite{dey2016OpticalCoherenceAtomicMonolayer}.
    }
    \label{fig:FIG_temp_dep}
\end{figure}

Shown in Fig.~\ref{fig:FIG_temp_dep} are values of $\tilde{\gamma}_{\infty}$ over a range of temperatures. Also shown are experimental measurements \cite{dey2016OpticalCoherenceAtomicMonolayer} of the temperature-dependent contribution to the optical linewidth, which were obtained by subtracting the lowest-temperature linewidth from the measured values across temperatures, similarly to what was done in Ref.~\citenum{chan2023exciton}. As can be seen, our calculations produce asymptotic linewidths that capture well the experimentally-observed temperature-dependent trend. While it is reassuring to see this trend being reproduced by our approach, it is especially the level of \emph{absolute} agreement that is noteworthy. We should emphasize that this level of agreement is reached without making any adjustments when parametrizing our approach based on the \emph{ab initio} values presented in the literature (cf.~Tab.~\ref{tab:parms}).

\section{Conclusions and outlook}\label{sec:conclusions}

In conclusion, we have presented a MQC framework for the microscopic and non-Markovian modeling of exciton--phonon scattering in solids, and applied it to model the optical linewidths of monolayer MoS$_2$. Through the application of BZ truncations, and by a systematic variation of the associated truncation parameters, we were able to obtain optical linewidths in the limit of an asymptotically-large material. To the best of our knowledge, our study is the first in exploring exciton--phonon scattering behaviors in this asymptotic limit. The asymptotic linewidths obtained through our approach are found to agree favorably with experimental measurements across a range of temperatures. On one hand, this level of agreement may not be entirely surprising, since all the model components included in our framework have previously been found to produce desirable results. On the other hand, the combination of these components has, again to the best of our knowledge, not been realized before. The accuracy reached within this combination is therefore reassuring.

Our results help establishing MQC dynamics as a viable and attractive tool for modeling carrier--phonon interactions in solids. Notably, the electronic and phononic truncations applied in our modeling, necessary for exploring the asymptotic limit, are uniquely afforded by the reciprocal-space MQC formalism that we previously introduced \cite{krotz2021reciprocal, krotz2022reciprocal}. In this regard, it is worth noting that we recently developed an analogous formalism, but expressed in arbitrary bases, by subjecting the complex-valued coordinates $\bm{z}$ to arbitrary unitary transformations \cite{miyazaki2024unitary}. This opens opportunities to study carrier--phonon interactions in the presence of material defects \cite{miyazaki2024unitary}.
    
Within our computational resources, we were able to reach sampling resolutions of the electron and hole BZ of up to $N_{\mathrm{k}} = 87$, albeit at slightly compromised values of the truncation radii $R_{\mathrm{k}}$ and $R_{\mathrm{X}}$. This amounts to a total of 7569 unit cells, which, as far as we are aware, exceeds system sizes so far explored in dynamics studies. The optical linewidth obtained in this calculation amounts to $\sim$6 meV at 300 K, which is still significantly lower than the asymptotic value of $\sim$10 meV. Although no quantitative conclusions can thus be drawn from this particular calculation, we expect it to still provide an unprecedented level of detail into the interactions between electron, hole, and phonons, offering an opportunity to study the resulting dynamics qualitatively, which we plan to further explore in future studies. In this regard, it will be particularly interesting to consider exciton dynamics beyond the line broadening mechanisms explored here, such as inter-valley scattering. Notably, Kelly and coworkers have recently reported on the simulations of inter-valley scattering in monolayer hexagonal boron nitride based on a mean-field MQC method similar to the approach presented here, although without invoking BZ truncations \cite{lively2024revealing}.

For longer-time dynamics, it may be necessary to resort to alternatives to mean-field MQC dynamics, due to the tendency of this method to overthermalize \cite{parandekar2005mixed, parandekar2006detailed}. A particularly-attractive alternative is provided by fewest-switches surface hopping \cite{tully1990molecular}, which we recently explored within a reciprocal \cite{krotz2022reciprocal} and arbitrarily-transformed \cite{miyazaki2024unitary} MQC framework. A recently-proposed coherent generalization of this method \cite{tempelaar2018generalization, bondarenko2023overcoming} may further enhance the accuracy through its improved ability to described the coherent dynamics prevalent in materials. Altogether, these developments provide new and exciting opportunities for the modeling of carrier--phonon dynamics in solids.

\section*{Acknowledgement}

The authors thank Denis Karaiskaj for providing the experimental data shown in Fig.~\ref{fig:FIG_temp_dep}. R.T.~thanks Andr\'{e}s Montoya-Castillo for helpful discussions. This material is based upon work supported by the National Science Foundation under Grant No.~2145433. 

\bibliography{bibliography}

\end{document}